\documentclass[a4paper]{spie}  

\usepackage{amsmath,amsfonts,amssymb}
\usepackage{graphicx}
\usepackage[colorlinks=true, allcolors=blue]{hyperref}
\usepackage{lipsum}
\usepackage{multirow,multicol}

\title{EuPRAXIA@SPARC\_LAB Status Update}

\author[a]{Fabio Villa}
\author[a]{David Alesini}
\author[a]{Maria P. Anania}
\author[a]{Marco Angelucci}
\author[b]{Alberto Bacci}
\author[a]{Antonella Balerna}
\author[a]{Marco Bellaveglia}
\author[a]{Angelo Biagioni}
\author[a]{Bruno Buonuomo}
\author[a]{Sergio Cantarella}
\author[a]{Fabio Cardelli}
\author[c]{Martina Carillo}
\author[d]{Mariano Carpanese}
\author[a]{Michele Castellano}
\author[c]{Enrica Chiadroni}
\author[e]{Alessandro Cianchi}
\author[a]{Fara Cioeta}
\author[b]{Marcello Rossetti Conti}
\author[a,f]{Marcello Coreno}
\author[c]{Lucio Crincoli}
\author[a]{Gemma Costa}
\author[a]{Alessandro Curcio}
\author[d]{Andrea Doria}
\author[a]{Alessio Del Dotto}
\author[a]{Mario Del Franco}
\author[a]{Martina Del Giorno}
\author[g]{Simone Di Mitri}
\author[a]{Enrico Di Pasquale}
\author[a]{Giampiero Di Pirro}
\author[a]{Alessandro Drago}
\author[a]{Zeinab Ebrahimpour}
\author[a]{Adolfo Esposito}
\author[a]{Luigi Faillace}
\author[a]{Antonio Falone}
\author[a]{Massimo Ferrario}
\author[c]{Luca Ficcadenti}
\author[a]{Giovanni Franzini}
\author[e]{Mario Galletti}
\author[a]{Alessandro Gallo}
\author[a]{Andrea Ghigo}
\author[a,g]{Luca Giannessi}
\author[a]{Anna Giribono}
\author[a]{Simona Incremona}
\author[c]{Pasqualina Iovine}
\author[a]{Franco Iungo}
\author[a]{Stefano Lauciani}
\author[a]{Andrea Liedl}
\author[a]{Valerio Lollo}
\author[c]{Stefano Lupi}
\author[a,f,h]{Augusto Marcelli}
\author[c]{Andrea Mostacci}
\author[d]{Federico Nguyen}
\author[a]{Michele Opromolla}
\author[e]{Gianmarco Parise}
\author[a]{Luigi Pellegrino}
\author[d]{Alberto Petralia}
\author[b]{Vittoria Petrillo}
\author[a]{Luca Piersanti}
\author[a]{Stefano Pioli}
\author[a]{Riccardo Pompili}
\author[g]{Emiliano Principi}
\author[a]{Ruggero Ricci}
\author[a]{Stefano Romeo}
\author[b]{Andrea R. Rossi}
\author[a]{Ugo Rotundo}
\author[a]{Lucia Sabbatini}
\author[i]{Andrea Selce}
\author[a]{Luisa Spallino}
\author[a]{Bruno Spataro}
\author[c]{Gilles J. Silvi}
\author[a]{Alessandro Stecchi}
\author[a]{Angelo Stella}
\author[e]{Francesco Stellato}
\author[a]{Federica Stocchi}
\author[a]{Cristina Vaccarezza}
\author[a]{Alessandro Vannozzi}
\author[a]{Sandro Vescovi}

\affil[a]{INFN Laboratori Nazionali di Frascati, Via E. Fermi 54, 00044 Frascati, Italy}
\affil[b]{Università degli Studi di Milano and INFN Sezione di Milano, Via Celoria 16, 20133 Milano, Italy}
\affil[c]{Università degli studi di Roma "Sapienza", P.le Aldo Moro 2, 00185 Rome, Italy}
\affil[d]{ENEA Frascati, Via E. Fermi 45, 00044 Frascati, Italy}
\affil[e]{University of Rome Tor Vergata, and INFN Sezione di Roma Tor Vergata, Via della Ricerca Scientifica 1, 00133 Roma, Italy}
\affil[f]{CNR Istituto Struttura della Materia and Elettra Sincrotrone Trieste, Basovizza Area Science Park, 34149 Trieste, Italy}
\affil[g]{Elettra Sincrotrone Trieste, Basovizza Area Science Park, 34149 Trieste, Italy}
\affil[h]{International Centre for Material Science Superstripes, RICMASS, Via dei Sabelli 119A, 00185 Rome, Italy}
\affil[i]{Università Roma Tre and INFN Sezione Roma Tre, Via della vasca navale 84, 00146 ROMA, Italy}

\authorinfo{Further author information: (Send correspondence to F.V.)\\ E-mail: fabio.villa@lnf.infn.it, Telephone: (+39) 0694032347}

\begin{document} 
\maketitle

\begin{abstract}
EuPRAXIA@SPARC\_LAB is a new multi-disciplinary user-facility that is currently under construction at the Laboratori Nazionali di Frascati of the INFN in the framework of the EuPRAXIA collaboration. The electron beam will be accelerated by an X-band normal conducting linac followed by a Plasma WakeField Acceleration (PWFA) stage. It will be characterized by a small footprint and it will drive two FEL beamlines for experiments, one in the VUV (50-180 nm) and the other in the XUV-soft x-rays (4-10 nm) spectral region. As an ancillary beamline, we are also including a betatron source in the x-ray from laser-plasma interaction. We present the status update of our facility.

\end{abstract}

\keywords{EuPRAXIA@SPARC\_LAB, FEL, Plasma acceleration}

\section{Introduction}


The introduction of Free Electron Lasers (FELs) has paved the way for an unprecedented range of experiments that capitalize on the distinctive properties of these radiation sources \cite{RevModPhys.88.015007, reviewer2, ackermann2007operation, emma2010first, huang2012sacla, allaria2015fermi, ko2017construction}. These sources boast high peaks brilliance, greater than $10^{30}\, $ $photons \, $ $s^{-1} \, $ $mrad^{-2} \, $ $mm^{-2} \, $ $0.1\%\, bandwidth$, short pulse duration in the range of tens of femtoseconds or less, and tunability in the VUV - X-ray energy range. They provide measurements with ultrafast time resolution and a high signal-to-noise ratio \cite{reviewer3, reviewer4, reviewer5}. As a result, many X-ray FEL facilities are currently operational, and there is a rising number of new facilities in the planning or active construction phase.

However, one significant drawbacks of present X-ray FEL facilities is the large amount of space required for electron accelerators, undulators, and photon beamlines. This typically involves hundreds of meters to a few kilometers, limiting FEL realization in large-scale laboratories.

To overcome this limitation, plasma wakefield acceleration (PWFA) is considered a promising technique for significantly reducing the required space for beam acceleration \cite{reviewer6, PhysRevA.44.R6189, leemans2006gev, blumenfeld2007energy, rosenzweig2010plasma, litos2014high, romeo2018simulation}. The high accelerating gradient of PWFA, which exceeds 1 GV/m, allows the facility's overall dimensions and cost to be reduced.

The EuPRAXIA Design Study aims to realize a FEL facility driven by plasma acceleration. The Frascati National Laboratories are proposing to host a new facility called $EuPRAXIA@SPARC\_LAB$ \cite{FERRARIO2018134},as part of the EuPRAXIA project which can meet these requirements. A high-brightness X-band linac, a plasma acceleration stage, and a FEL comprise this facility. 

The facility will also have an ultrafast high power laser, as an upgrade of our FLAME laser system \cite{bisesto2018flame}, capable to operate both independently from the accelerator to produce PWFA beams from self-injection schemes \cite{costa2018characterization} or with the accelerator as a driver for PWFA \cite{ROSSI201460}. It is possible to have a betatron source in the x-ray spectrum using similar configurations to PWFA\cite{CURCIO2017388}. 

In this proceeding, we will discuss the status of the machine for the two main FEL beamlines and, in the last section, of the ancillary light sources we are planning to add to the facility. 

\subsection{EuPRAXIA@SPARC\_LAB layout} 

\begin{figure} [ht]
  \begin{center}
    \begin{tabular}{c} 
       \includegraphics[height=5cm]{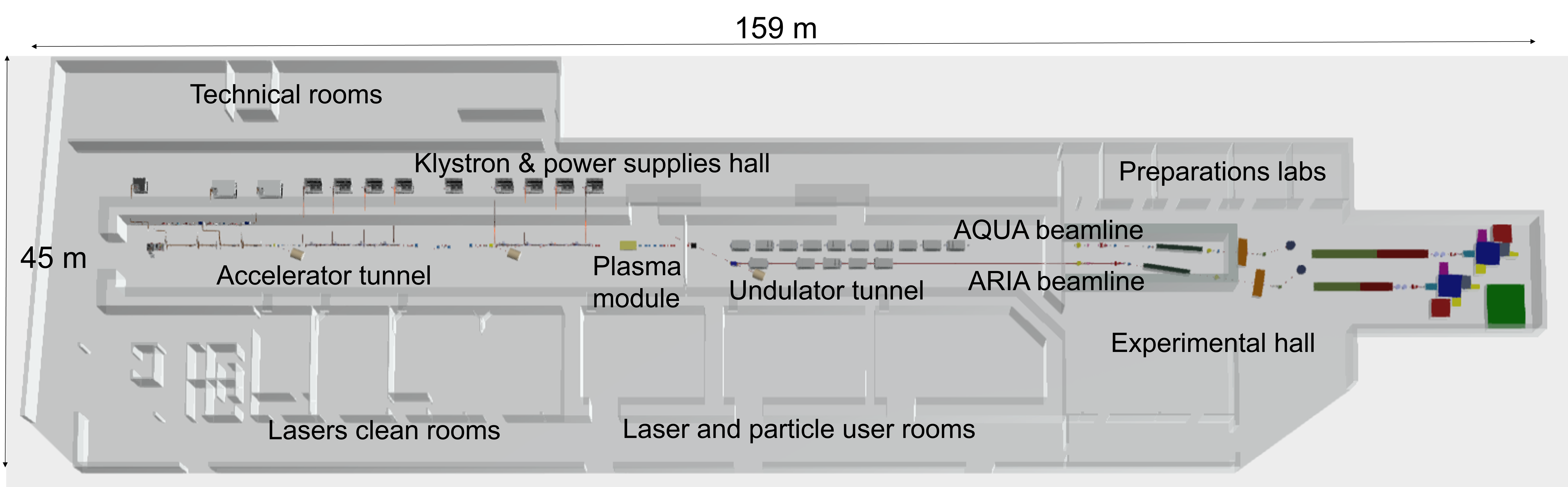}
	\end{tabular}
	\end{center}
  \caption[ESL-layout-img] 
  { EuPRAXIA@SPARC\_LAB layout
}\label{ESL-layout-img}
   \end{figure}

Fig. \ref{ESL-layout-img} depicts the general layout of the EuPRAXIA@SPARC LAB. Rooms for cooling, building air conditioning, data storage, klystrons, power supplies, and all machine services are located at the top of the building. The accelerator and undulators are housed in a shielded tunnel with 2 m thick concrete walls in the center. An experimental hall on the right of the shielded tunnel resides the FEL beamlines as well as the ancillary labs needed for sample and component preparation. There are clean rooms for the laser systems and a shielded section for experiments with electrons and particles, accelerated by a laser or by the linac, in the bottom part of the building layout. The laser power supplies, control rooms, and some offices will be located on a second floor that hasn't been shown in the figure. 

\section{Electron accelerator}

\begin{figure} [ht]\label{ESL-acc-img}
  \begin{center}
    \begin{tabular}{c} 
       \includegraphics[width=0.95\textwidth]{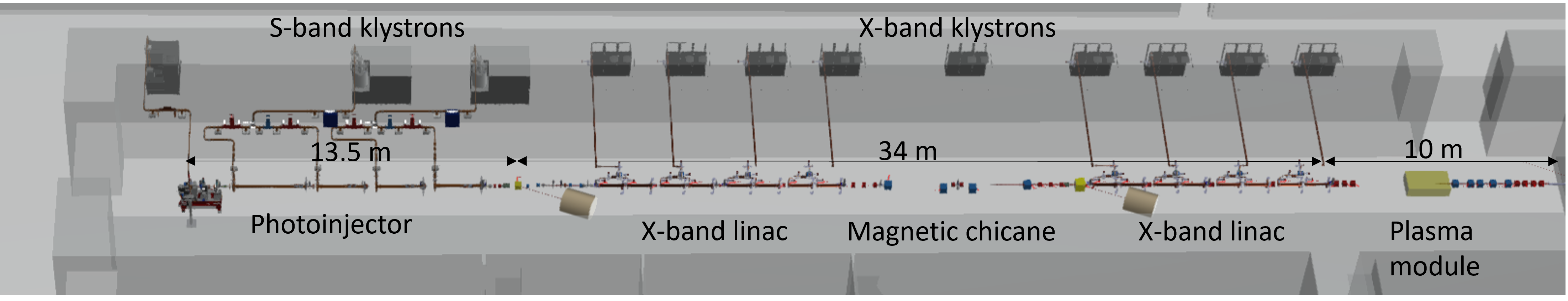}
	\end{tabular}
	\end{center}
  \caption[ESL linac img] 
  { EuPRAXIA@SPARC\_LAB accelerator layout
}
   \end{figure}

The detailed layout of the accelerator is shown in Fig. \ref{ESL-acc-img}. The accelerator is composed of a conventional RF linac followed by an additional acceleration stage based on PWFA.

\subsection{RF accelerator}

The RF accelerator consists of a photoinjector, which generates the high-brilliance electron beam, and an X-band linac, which is used to achieve high energy.

The copper photocathode of the injector is illuminated by a Ti:Sapphire femtosecond laser pulse to generate photoelectrons, which are then accelerated and focused into a high-quality electron beam in the S-band RF gun.
The gun can be operated in single-bunch mode, where a single bunch of electrons up to 500 pC is generated and accelerated, or in multi-bunch mode, where multiple laser pulses are sent with ps-range separation to have multiple bunches of electrons generated and accelerated in the same RF bucket \cite{villa2016laser} The multi-bunch option can be tuned to provide different bunch spacing, number of bunches, and other parameters. The tunable multibunch option for this photoinjector gun is needed to optimize the PWFA section. After the RF gun, there is a X-band linarizer \cite{ficcadenti2019xls} to optimize the longitudinal phase-space of the electrons. 4 S-band RF sections, the first 3 meters long and the others 2 meters long, will accelerate the electrons up to 120 MeV. The first RS section is used in a velocity bunching configuration \cite{ferrario2010experimental} to compress the beam. 

The beam is then accelerated by two sets of 8 X-band RF linac sections \cite{diomede2018preliminary}, separated by a magnetic chicane to further compress the electrons. The beam energy at the linac end is up to about 500 MeV when we use the following PWFA section for better control of the electron bunches parameters. In the case of a longer bunch that won't exploit the PWFA section, the beam energy of the single bunch configuration can be raised up to 1 GeV for best FEL performances (this working point is reported in tab \ref{tab:ele_WPs} in the X-band columns). The RF accelerator can work up to the X-band klystron repetition rate; state-of-the-art klystrons can reach 100 Hz, while new ones are under study for reaching up to 400 Hz repetition rate. 


\begin{table}[ht]
\caption{Working points for electron beams in the AQUA and ARIA beamlines. For each beamline, there is a working point that uses PWFA electrons and another that uses only the X-band linac at full power} 
\label{tab:ele_WPs}
\begin{center}       
\begin{tabular}{|l|c|c|c|c|} 
\hline
 \rule[-1ex]{0pt}{3.5ex} Electron parameters & \multicolumn{2}{|c|}{AQUA} & \multicolumn{2}{|c|}{ARIA}  \\
 \rule[-1ex]{0pt}{3.5ex} & PWFA & X-band & PWFA & X-band \\
 \hline
 \rule[-1ex]{0pt}{3.5ex} Charge (pC) & 30 & 200 & 30 & 200  \\
\hline
 \rule[-1ex]{0pt}{3.5ex} Peak current (kA) & 1.8 & 1.8 & 1.8 & 0.8 \\
\hline
 \rule[-1ex]{0pt}{3.5ex} Energy (Max range, GeV) & 1.2 & 1.0 & 1.2 & 1.0  \\
\hline
 \rule[-1ex]{0pt}{3.5ex} Normalized emittance (slice, mm-mrad) & 0.6  & 0.5 & 0.8 & 1.5  \\
\hline
 \rule[-1ex]{0pt}{3.5ex} Energy spread (slice, \%) & 0.022 & 0.05 & 0.022 & 0.02  \\
\hline
 \rule[-1ex]{0pt}{3.5ex} Repetition rate (Hz) & 1-10  & 100 & 1-10 & 100  \\
\hline

\end{tabular}
\end{center}
\end{table}

\subsection{PWFA Section}

\begin{figure} [ht]
  \begin{center}
    \begin{tabular}{c} 
       \includegraphics[width=0.95\textwidth]{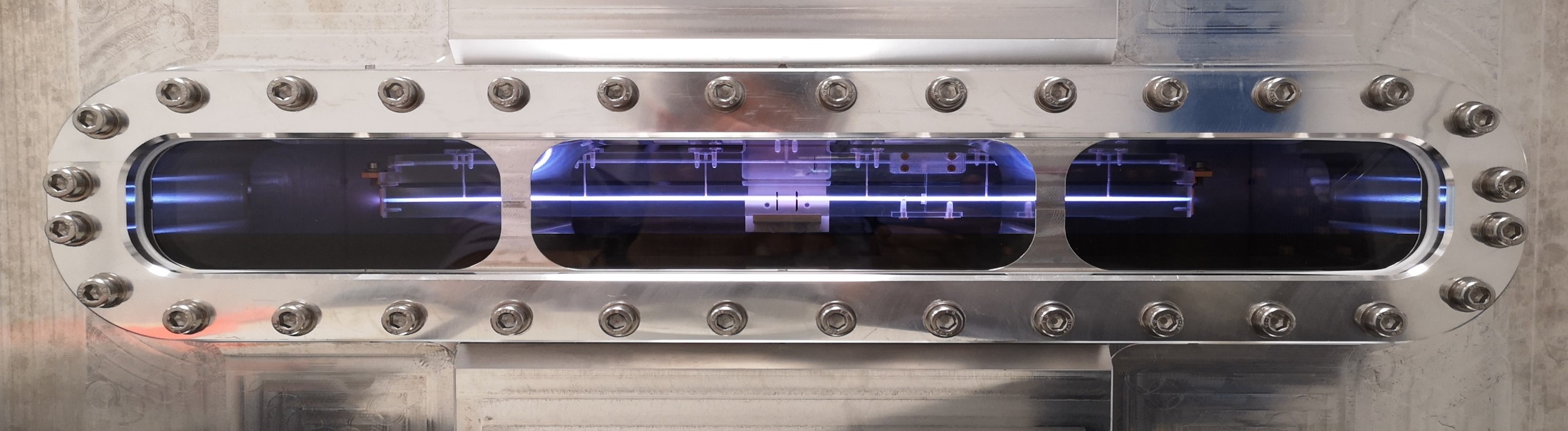}
	\end{tabular}
	\end{center}
  \caption[capillary] 
  { Discharge in a 40 cm capillary in SPARC\_LAB. Full characterization of the plasma density, vacuum management and ageing is ongoing.
}\label{capillary}
   \end{figure}

Plasma acceleration is capable of achieving high acceleration gradients (in the order of 1 GeV/m), reducing the dimension of the accelerator. Different configurations of drivers (using laser pulses \cite{wang2021free, labat2023seeded} or electron bunches \cite{pompili2022free, galletti2022stable}) can  accelerate electron beams capable of driving a FEL exponential growth. Following our expertise in SPARC\_LAB, we plan to use a particle driven PWFA in a plasma capillary. The PIC simulations we are doing show that a 200 pC bunch can drive the acceleration of the subsequent 30 pC bunch to more than 1 GeV and with good electron parameters (reported in tab \ref{tab:ele_WPs} in the PWFA columns) in a about 50 cm capillary. Studies on single long capillary (fig. \ref{capillary}) and staging of smaller ones are now undergoing in SPARC\_LAB \cite{costa2022characterisation}. 

The repetition rate of the PWFA section is limited by the time required to remove the spent gas from the chamber and restore high vacuum in the chamber. We have currently achieved a repetition rate of 1 Hz, with 10 Hz appearing to be feasible in the near future. In our longer-term plan, we intend to improve this repetition rate to be matched with that of the linac. 

The plasma module will include dedicated plasma diagnostics\cite{arjmand2020characterization} to characterize the process and optimize the PWFA parameters. After the capillary, the spent driver must be discarded because its energy will be lower than the accelerated one driving the FEL but with enough particles and energy to damage the first undulators with many shots. After the plasma section, quadrupoles or active plasma lenses\cite{pompili2019plasma} in a focusing-defocusing configuration will be inserted with small apertures in the focal position of the FEL driving bunch to remove the spent PWFA driver bunch in subsequent stages. The electron beam will be injected into the two FEL beamlines via a small chicane and matching quadrupoles.

Star-to-end simulation of the working points is ongoing, optimizing the parameters of the accelerators to have better FEL performances. Reference working points are reported in table \ref{tab:ele_WPs}, while we are now studying the parameter stability and the jitters of the working point.

\section{FEL beamlines}

\begin{figure} [ht]
  \begin{center}
    \begin{tabular}{c} 
       \includegraphics[width=0.95\textwidth
    ]{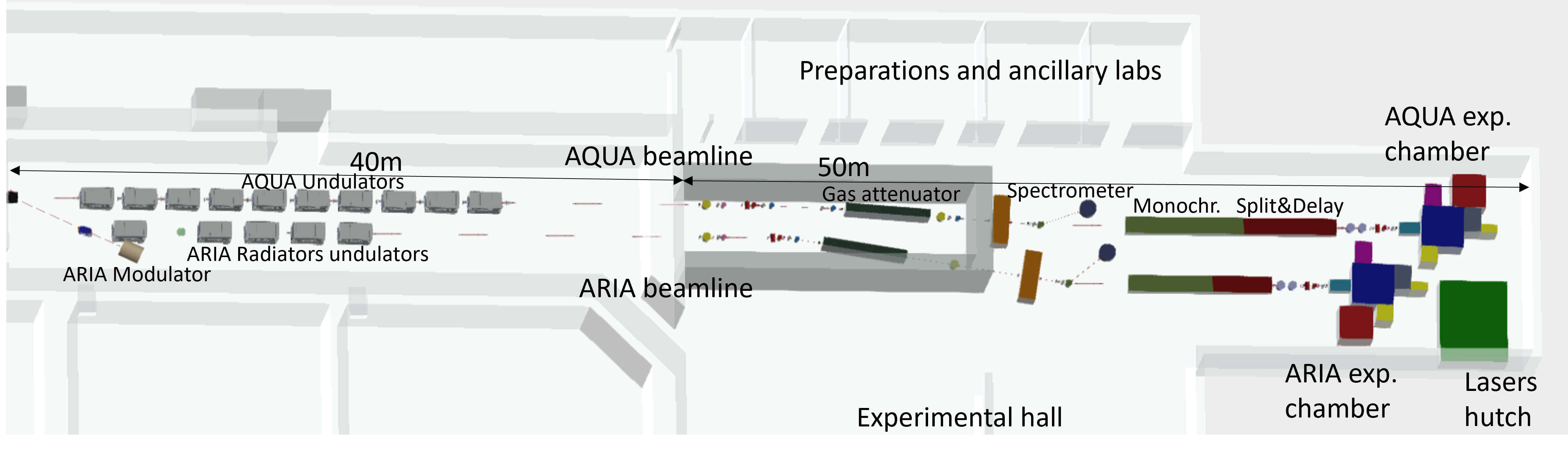}
	\end{tabular}
	\end{center}
  \caption[ESL beamlines img] 
  { EuPRAXIA@SPARC\_LAB FEL beamlines AQUA and ARIA layout
}\label{ESL beamlines img}
   \end{figure}

Two FEL beamlines with different undulator configurations are foreseen in the project. The top one in fig. \ref{ESL beamlines img} is called AQUA and will operate in the water window in SASE configuration \cite{sase}, the lower one, called ARIA, will operate in the VUV range in seeding configuration.

\subsection{AQUA beamline}

The photon wavelengths of the AQUA beamline will reach the so-called "water window" in the wavelength range between carbon (4.40 nm) and oxygen (2.33 nm) K-absorption edges. Since biological samples are mainly composed of light atoms (mostly carbon) and find their native environment in aqueous solutions, the absorption contrast between the carbon atoms of the biological samples and the oxygen of the water surrounding them is the highest in the water window. As a result, measurements of unstained cells, viruses, and organelles in their hydrated, native state have become possible \cite{nakasako2020methods,gallagher2016frontier}.  Aside from photon spectroscopy and imaging experiments, the AQUA beamline will also allow for resonant inelastic X-ray scattering, electron spectroscopies, and photo-fragmentation measurements \cite{principi2020atomic,ebrahimpour2022progress}. 

In order to reach the water window, a GeV electron beam (either from PWFA or from the X-band RF linac at full power) is sent to ten Apple-X undulators with a period of 18 mm. Those undulators will produce SASE FEL light, with typical parameters (simulated with Genesis \cite{reiche1999genesis}) reported in tab. \ref{tab:AQUA_BL}. The undulator can also change the beam polarization, for example, from linear to circular which allows the photoelectron circular dichroism (PECD) to study chirality effects in molecules \cite{rouxel2022molecular}. Although the energy in our machine is limited to 1 GeV with RF linac, we can reach higher energies with PWFA and, in the future, go below the 4 nm of our current start-to-end simulation from the cathode to the FEL. 

\begin{table}[ht]
\caption{The undulator and FEL radiation parameters of the AQUA beamline at the undulator's end.} 
\label{tab:AQUA_BL}
\begin{center}       
\begin{tabular}{|l|c|c|} 
\hline
 \rule[-1ex]{0pt}{3.5ex} Undulator parameters & \multicolumn{2}{|c|}{AQUA}   \\
\hline
 \rule[-1ex]{0pt}{3.5ex} Period (mm) & \multicolumn{2}{|c|}{18}   \\
\hline
 \rule[-1ex]{0pt}{3.5ex} Max strength (k) & \multicolumn{2}{|c|}{1.47}   \\
\hline
 \rule[-1ex]{0pt}{3.5ex} Min gap (mm) & \multicolumn{2}{|c|}{6}   \\
\hline
 \rule[-1ex]{0pt}{3.5ex} Active length (m) & \multicolumn{2}{|c|}{19.8}   \\
\hline
\hline
 \rule[-1ex]{0pt}{3.5ex} Radiation parameters & PWFA & X-band   \\
\hline
 \rule[-1ex]{0pt}{3.5ex} Energy per pulse ($\mu$J) & 10 & 10  \\
\hline
 \rule[-1ex]{0pt}{3.5ex} Wavelength tunability (nm) & 4-10 & 4-10  \\
\hline 
 \rule[-1ex]{0pt}{3.5ex} Bandwidth (\%) & 0.3 & 0.3  \\
\hline 
 \rule[-1ex]{0pt}{3.5ex} Pulse length (fs) & 15 & 60  \\
\hline

\end{tabular}
\end{center}
\end{table} 

 The undulator's light is then fully characterized before being delivered to the experiments via the photon beamline \cite{villa2020photon}. The beamline transmission is about 34\%,  due to primarily the Ni-coated mirror reflectivity (in the order of 80\% per mirror at 2$^\circ$ grazing angle). There is the possibility to insert in the beamline a split-and-delay system to have two geometrically separated beams in the picoseconds range, with an additional loss of about 60\% of the beam due to mirror reflectivity. Another system needed by some experiments that can be inserted on the beam is a grating monochromator. A grating monochromator is another system required by some experiments that can be inserted on the beam. Due to the short temporal length of the beam, a compensating grating configuration will be used \cite{poletto2014double} to avoid chirping the beam, but with a beam transmission of about 3\% due to grating diffraction efficiency and mirror reflectivity, in addition to beamline losses and the amount of spectrum cut by the monochromator. The stability issues in the monochromator caused by the SASE process and PWFA acceleration will be investigated further. A gas intensity monitor \cite{tiedtke2008gas}, removable scintillating screens, an online grating spectrometer, and beam position monitors will be among the beam diagnostics. Based on gas ionization streaking, dedicated instrumentation for full temporal characterization and polarization measurements will be investigated \cite{heider2019megahertz}. The beam will be focused at the end by K-B mirrors inside the experimental chamber, which will be outfitted with the necessary instrumentation and sample delivery \cite{condmat4010030}.

\subsection{ARIA beamline}

The second FEL beamline we are planning is called ARIA, and it will operate at a lower wavelength energy in the VUV range (around 50-180 nm), is easier to realize, has a larger input parameter acceptance, and requires fewer undulators to operate (thus  having a lower economic impact on the project); for those reasons it can begin operations sooner than the AQUA beamline and it can also be used for machine commissioning.
The seeded VUV ARIA FEL line exploits the standard High-Gain Harmonic Generation (HGHG) configuration \cite{yu2000high, yu2003first, togashi2011extreme} with one Apple-II modulator undulator and four radiator undulators. A commercial OPG-OPA Ti:Sapphire laser system will generate the seed pulse, with a tunability in the range of 600-800 nm (for the fundamental) and 320-400 nm (for the second harmonic), an energy per pulse of more than 20 $\mu$J in all the spectral range, and a time duration of about 200 fs. Typical parameters (simulated with Genesis) are reported in Tab. \ref{tab:ARIA_BL}, where the undulator will exploit the short beam duration of the PWFA electrons to have very short pulses or the capability of a larger charge of the X-band only option to reduce the pulse bandwidth. The simulated number of photons in both cases will be almost the same. 

\begin{table}[ht]
\caption{The undulator and FEL radiation parameters of the ARIA beamline at the undulator's end.} 
\label{tab:ARIA_BL}
\begin{center}       
\begin{tabular}{|l|c|c|} 
\hline
\rule[-1ex]{0pt}{3.5ex} Undulator parameters & \multicolumn{2}{|c|}{ARIA}   \\
\rule[-1ex]{0pt}{3.5ex}   & modulator & radiator\\
\hline
\rule[-1ex]{0pt}{3.5ex} Period (mm) & 100 & 55   \\
\hline
\rule[-1ex]{0pt}{3.5ex} Active length (m) &  3.0 & 8.4  \\
\hline
\rule[-1ex]{0pt}{3.5ex} Seeding wavelengths (nm) & \multicolumn{2}{|c|}{320-400 + 600-800}\\
\hline
\rule[-1ex]{0pt}{3.5ex} Seeding energy per pulse ($\mu$J) & \multicolumn{2}{|c|}{$>20$}\\
\hline
\rule[-1ex]{0pt}{3.5ex} Seeding length (fs) & \multicolumn{2}{|c|}{200}\\
\hline
\hline
\rule[-1ex]{0pt}{3.5ex} Radiation parameters & PWFA & X-band   \\
\hline
\rule[-1ex]{0pt}{3.5ex} Energy per pulse ($\mu$J) & 200 & 200  \\
\hline
\rule[-1ex]{0pt}{3.5ex} Wavelength tunability (nm) & 50-180 & 50-180  \\
\hline 
\rule[-1ex]{0pt}{3.5ex} Bandwidth (\%) & 3 & 0.05  \\
\hline 
\rule[-1ex]{0pt}{3.5ex} Pulse length (fs) & 15 & 100  \\
\hline

\end{tabular}
\end{center}
\end{table} 

The ARIA photon beamline will be conceptually similar to the AQUA beamline, but due to the different energy ranges of the photons with different coatings and materials of the beam instrumentation, such as mirrors, filters, and gratings \cite{villa2022aria}. We expect higher overall transmission than the one in the AQUA beamline due to the higher reflectivity of the C-coated optics in the VUV range, the optics are now under study. Removable split-and-delay and monochromator systems will be available in the beamline.

The ARIA beamline can support a wide range of experiments in atomic, molecular, and cluster physics, as well as solid, liquid, and gas phase materials\cite{villa2022aria}. The ARIA energy range allows us to probe new electronic transitions well within the 7-20 eV range for classes of cluster materials such as nano-carbons \cite{bogana2005leaving} and potential gap dielectrics such as metal oxides using the ultra-fast pump-probe configuration. Photoemission spectroscopy, photo-electron circular dichroism, photo-fragmentation of molecules, and time of flight are examples of experimental techniques we intend to construct that provide powerful tools for investigating excited states such as excitons, polarons, and spin-charge-orbital ordering \cite{hofer1997time}. Using this experimental techniques, various materials and compositions can be studied with potential applications in hybrid organic–inorganic hetero-junctions of organic solar cells \cite{huang2008electronic}, as well as in modulation-doped semiconductor heterostructures of metal–semiconductor interfaces and spintronics devices such as spin valves or diluted magnetic semiconductors, among others.

\section{Other light sources}

In addition to the above-mentioned FELs, other light sources will be available at EuPRAXIA@SPARC\_LAB, including a 500 TW high-power laser system, whose front-end will be shared with the photocathode and linac diagnostics. Commercially available laser systems of such power have a repetition rate of up to 5 Hz; we anticipate systems with a repetition rate of up to 10 Hz will be available in the future. Depending on the laser and target configuration, this laser system can generate plasma accelerated particles (electron \cite{tajima1979laser, costa2018characterization} or protons \cite{clark2000measurements, bisesto2019single} and ions). Along with the access to the laser clean room, the laboratory will have a shielded, dedicated area for those experiments. Additional experimental configurations employing laser-generated particles and linac-accelerated electrons are being examined. The same laser can also be used to operate a laser-driven PWFA \cite{bisesto2016laser} of an electron bunch originating from the RF linac.

The high-power laser can also be used as a betatron source to generate x-ray bursts \cite{rousse2004production, CURCIO2017388}. Betatron radiation is the wiggler-like synchrotron light produced by electrons accelerated in plasma wakefields when electrons are injected off-axis, where intense transverse forces present. The laser pulse will drive the plasma wakefield, and electrons from the plasma are self-injected in the region of strong fields, resulting in a source that is only a few meters long, including the space for laser injection and extraction. The emitted radiation is a "white beam" with a very short duration, in the same order of the laser pulse, and with a continuous spectral distribution similar to synchrotron radiation with a cut-off frequency that varies depending on the laser and plasma characteristics (typically in the keV to tens of keV range). This source can be used solely or in combination with a linac electron or FEL light in a pump-probe configuration. This source could be used for imaging and X-ray absorption spectrometry  \cite{stellato2022plasma}, for example. We recently launched the EuAPS ("EuPRAXIA Advanced Photon Source") initiative, which will be funded in the framework of "Next Generation EU" (by the Italian "Piano Nazionale di Ripresa e Resilienza", PNRR) to host such a source in SPARC\_LAB. The development and expertise acquired through this project will be included in the EuPRAXIA@SPARC\_LAB facility. 

THz beam can also be generated through laser down-conversion \cite{kitaeva2008terahertz} or linac-based employs  \cite{chiadroni2020versatile}. The possible applications of the THz source are under study and dedicated space will be available in the laser clean rooms and in the experimental hall. 

\section{Conclusions}

We presented the progress of our planned FEL facility EuPRAXIA@SPARC\_LAB. The technical design report will be published by the end of 2025. Meanwhile, we are moving forward with the civil permits for the building's realization; we plan to begin construction in 2024 and complete it by 2027. The machine's installation and operation will begin as soon as the structure is ready to receive it and will be commissioned by 2028. If no unforeseen delays happen, we anticipate having pilot users in 2029.



\bibliography{report} 
\bibliographystyle{spiebib} 

\end{document}